# New Approach to Determine the Quality of Graphene

Soo-Whan Kim[1], Hyun-Kyung Kim[2,a)], Sunghun Lee[1], Kyujoon Lee[3], Joong Tark Han[4], Kwang-Bum Kim[2], Kwang Chul Roh[5], Myung-Hwa Jung[1,d]

[1] *Department of Physics, Sogang University, Seoul 121-742, South Korea.*

[2] *Department of Material Science and Engineering, Yonsei University, Seodaemun-Gu, Seoul 120-749, South Korea*

[3] *Institut für Physik, Johannes Gutenberg-Universität Mainz, Staudinger Weg 9, 55128 Mainz, Germany*

[4] *Nano Carbon Materials Research Group, Creative and Fundamental Research Division, Korea Electrotechnology Research Institute, 12, Bulmosan-ro 10beon-gil, Seongsan-gu, Changwon 642-120, South Korea*

[5] *Energy Efficient Materials Team, Energy & Environmental Division, Korea Institute of Ceramic Engineering & Technology, Jinju 660-031, South Korea*

The reduction of graphene oxide is one of the most facile methods to fabricate a large amount of graphene and the reduction rate of graphene oxide is related with the quality of synthesized graphene for its possible application. The reduction rate is usually determined by using various spectroscopy measurements such as Raman spectroscopy, Fourier transform infrared spectroscopy, and X-ray photoelectron spectroscopy. Here we propose that the magnetic data can be used as a means of determining the quality of graphene oxide (GO) and reduced graphene oxide (RGO) by the investigation of close relation between magnetic moment and chemical bonding state. Our experimental findings and previous theoretical studies suggest that hydroxyl functional groups in GO mainly contribute to Langevin paramagnetism, carboxyl functional groups in RGO1 act as the source for Pauli paramagnetism, and $sp^2$ bonding state in RGO2 plays a major role on the diamagnetism. Especially in terms of mass production, the magnetic data is useful for decomposing the chemical bonding electronic states in graphene-like samples and judging their quality.

---

a) Present address: Department of Material Science & Metallurgy, Cambridge University, Cambridge, CB3 0FS, UK

b) Author to whom correspondence should be addressed. Electronic mail: mhjung@sogang.ac.kr, Tel. +82-2-705-8828

Although a single graphene sheet, composed of $sp^2$-carbons arranged into two-dimensional fused hexagon, was initially thought to be thermodynamically unstable,[1] its large surface area to mass ratio, excellent electrical conductivity, and extremely high mechanical strength have led to tremendous researches in many current fields of science for last ten years.[2-5] In early stage, the exfoliation of bulk graphite is the most commonly used method for the production of graphene sheets.[6,7] However, such method has the limitation in terms of mass production, so that numerous research groups have developed the epitaxial growth methods based on chemical vapor deposition.[8-11] In parallel, to realize high-quality graphene on a large scale, the use of chemically modified graphene have been attempted, which enables the formation of graphene-like materials, namely graphene oxide (GO) and its reduced form (RGO).[12-17] The reduction of GO is an extremely important process because it has a significant impact on the quality of RGO, which determines how close RGO comes to pristine graphene in terms of structure. Moreover, since various kinds of oxygen-containing functional groups can remain during the reduction process of GO into RGO, one key challenge is the removal of functional groups and the restoration of $sp^2$ bonding. An ideal graphene should be nonmagnetic itself due to $\pi$-symmetry electronic states formed by the unhybridized $p_z$ atomic orbital of $sp^2$ carbon atoms.[18] When the defects like vacancies are introduced in lattice structure of graphene, magnetic moments in graphene could be introduced by removing the $p_z$ orbital from $\pi$-symmetry states.[19] These results indicate that the magnetic moment is strongly interconnected with the chemical bonding states, for example, oxygen-containing and/or $sp^2$-carbon bonding states.

Herein, we demonstrate for the first time that the magnetic data can be used as a means of determining the quality of RGO from the close relation between magnetic moment, oxygen-containing functional groups, $sp^2$ carbon bonding, and reduction rate of GO. This is compared with various spectroscopic methods used for the characterization of RGO such as Raman spectroscopy, Fourier transform infrared spectroscopy, and X-ray photoelectron spectroscopy. Based on the precise information on chemical states of GO and RGO obtained from spectroscopic studies, we analyze the magnetic data. Since the spectroscopic measurements are usually carried out in a local area and the magnetic measurements are performed with a bulky sample, the magnetic data can provide more useful information on the quality of RGO in bulk-scale production.

There are three important steps for graphene production; oxidation of graphite, exfoliation of graphite oxide, and reduction of graphene oxide. The extent of graphite oxidation is dependent on the synthetic method as well as the reaction time. In our case, commercial graphite is oxidized into graphite oxide by using two different methods of Hummers and Brodie,[20,21] and next graphite oxide is exfoliated into graphene oxide (GO) by sonicating or stirring in solutions such as diethylene glycol and aqueous NaOH. Finally, GO is reduced into graphene-like material by hydrothermal treatment or hydrazine solution.[22,23] One reduced GO (called RGO1) in this manuscript was fabricated by microwave-assisted hydrothermal treatments of GO prepared by Hummers method, which is identical method described in Ref. 22. The other reduced GO (called RGO2) is reduced from the GO oxidized via Brodie method by adding hydrazine, as previously synthesized in Ref. 23. The structural changes from GO to RGO were investigated by measuring X-ray diffraction (XRD, Cu K$\alpha$, Rigaku) and Raman spectroscopy (LabRam, JovinYvon). Fourier transform infrared spectroscopy (FT-IR, Vertex 70, Bruker) and X-ray photoelectron spectroscopy (XPS, Escalab 220i-XL, VG Scientific Instrument) were used to examine the changes in the chemical bonding of carbon and oxygen-containing functional groups after reduction process. The magnetic properties of as-synthesized GO and RGO samples were measured with a superconducting quantum interference device-vibrating sample magnetometer (SQUID-VSM) at temperatures from 2 and 300 K and in magnetic field up to 70 kOe.

Figure 1(a) shows the XRD patterns of GO, RGO1, and RGO2. The transformation of GO to RGO is revealed by the disappearance of the characteristic peak of GO at $2\theta = 9°$ and the appearance of the characteristic peak of graphene at $2\theta = 25.5°$. The main diffraction peak of GO at $2\theta = 9°$ gives an interlayer separation of 9.7 Å, which is significantly higher than that of graphite (3.35 Å).[24] It is attributed to the presence of a variety of oxygen-containing functional groups in GO.[25] In both RGO samples, the diffraction peak related with GO is not detected. The broad peak at $2\theta = 25.5°$ correspond to the interlayer spacing of 3.43 Å, indicating the presence of graphene and the removal of oxygen-containing functional groups.

As aforementioned, the oxygen reduction rate and its related chemical states are, in general, characterized by optical measurements. Raman spectra are collected in Fig. 1(b). According to the reported Raman spectrum of graphene,[26,27] the peak at 1585 cm$^{-1}$ corresponds to the E$_{2g}$ mode (G band) of graphite, which is associated with the vibration of *sp*$^2$-bonded carbon atoms in two-dimensional hexagonal lattices. The peak at 1350 cm$^{-1}$ is related with the vibration of carbon atoms with dangling bonds in the plane terminations of disordered graphite, which corresponds to the breathing mode (D band) of *k*-point phonons of A$_{1g}$ symmetry. The Raman spectrum of GO is similar to that of RGO1 and dissimilar to that of RGO2. There is no noticeable difference of relative peak height between G and D bands in both GO and RGO1, suggesting that defects such as vacancies generated during the oxidation of graphite remain unchanged even after the removal of oxygen-containing functional groups. For RGO2, on the other hand, the peak height of D band is much higher than that of G band, meaning the high quality of graphene structure. Additional peak of 2D band is also indicative of high quality of RGO2. The narrow and symmetric 2D-band peak centered at 2687 cm$^{-1}$ shows typical feature of monolayer or bilayer graphene.[27] It should be noted that the Raman spectrum of RGO2 is similar to the previous result of "supra-RGO" reported in Ref. 28. Two more peaks of D' and D+D' bands are also observed, implying that RGO2 is not as perfect as it should be, due to a small fraction of randomly distributed impurities or surface charges.[29] The quality of the GO, RGO1 and RGO2 samples will be discussed further below, especially in terms of the reduction rate.

FT-IR spectrum can be used as a measure that oxygen-containing functional groups are introduced during oxidation and these functional groups are decomposed after reduction process.[30] Figure 1(c) shows the comparative FT-IR spectra of GO, RGO1, and RGO2. The characteristic features of GO indicate the presence of oxygen-containing functional groups on the surface. Each IR peak in GO corresponds to carbonyl C=O (1733 cm$^{-1}$), aromatic C=C (1620 cm$^{-1}$), carboxy C–O (1413 cm$^{-1}$), epoxy C–O (1224 cm$^{-1}$), and alkoxy C–O (1051 cm$^{-1}$), respectively.[26] For RGO1, the oxygen-containing functional groups of carboxy C–O and alkoxy C–O vanish, suggesting the reduction of GO sheets, and additional peak appears at 1577 cm$^{-1}$, which is related with skeletal vibrations from un-oxidized graphitic domains.[26] However, the epoxy C-O group still remains after reduction, implying that the chemical reduction of GO prepared by Hummers method results in incompletely reduced products. For RGO2, most oxygen-containing functional groups disappear and only C=C group is detected, meaning that RGO2 is successfully reduced to graphene. These results are in accord with the Raman spectra, i.e., being indicative of high quality of RGO2.

XPS has been employed to quantitatively determine the elemental composition and the bonding nature of GO and RGO samples. Figure 1(d) shows wide-scan XPS spectra of all the samples. It is clear that the O1s peak is dominant for GO, while the C1s peaks are dominant for RGO1 and RGO2. This result indicates the removal of oxygen-containing groups in the RGO samples. The intensity ratio of C1s to O1s of RGO2 is higher than that of RGO1, meaning that RGO2 is more reduced, in consistent with the Raman spectroscopy and FT-IR results. The existence of N1s peak in RGO 2 may be attributed to the hydrazine used for the reduction process.[31]

Figure 2 presents the curve-fitted C1s spectra, which consist of five carbon-related bonding: C=C ($sp^2$, ~284.7 eV), C–C ($sp^3$, ~285.5 eV), C–O (hydroxyl, ~286.5 eV), C=O (carbonyl, ~288.3 eV), and O–C=O (carboxyl, ~289 eV) as reported previously.[32,33] The deconvoluted results of relative elemental composition percentage of $sp^2$, $sp^3$ and oxygen-containing functional groups for GO, RGO1 and RGO2 are summarized in Table 1. The contribution of heterocarbon components from oxygen-containing functional groups in GO is 58.30%, which is larger than that of carbon components from $sp^2$ and $sp^3$. This result implies that the oxygen-containing functional groups are strongly attached to GO surfaces by oxidation treatment. After the reduction, the contribution of hydroxyl and carbonyl groups considerably decreases. For both RGO samples, the ratio of $sp^2/sp^3$ and C/O increases dramatically, revealing that most oxygen-containing functional groups are removed after the reduction. They are further enhanced in RGO2, being indicative of more reduction in RGO2. As explained above, therefore, RGO2 seems to be the most similar graphene-like hexagonal lattice.

So far, we have analyzed the spectroscopy data obtained from Raman spectroscopy, FT-IR, and XPS as a means of determining the quality of RGO, especially, by examining the change of oxygen-containing functional groups and $sp^2$ carbon bonding after oxygen reduction of GO. Although the spectroscopy measurements are useful, there are some limitations for each measurement.[34] Raman spectroscopy has difficulty in providing specific information on the oxygen-containing functional groups because it is sensitive to homonuclear molecules.[35] The disadvantage of FT-IR is that some bonding such as C-C bonding has the weak or zero vibrational activity,[36] but it gives concise information on the change of oxygen-containing functional groups in RGO after reduction process. XPS is a surface sensitive technique that can be used to quantitatively probe the carbon and oxygen bonding. In addition, each spectroscopic result provides different information on the chemical states of graphene. These are the reason why the combined spectroscopic results are required for evaluating the quality of graphene. From now, we demonstrate for the first time that the magnetic data can be used to determine the quality of RGO from the close relation between magnetic moment, oxygen-containing functional groups, $sp^2$ carbon bonding, and reduction rate of GO. The magnetic data will be discussed in comparison with the spectroscopic results above. It should be noted that the magnetism is a bulk property, while the spectroscopy measurements are surface sensitive and are carried out in a local area. Thus, the magnetic data can provide more useful information on the quality of RGO in terms of mass production, although the spectroscopy data give the precise information on the chemical states of GO and RGO.

As aforementioned, conventional graphene obtained from the exfoliation of bulk graphite is known to be diamagnetic due to a delocalized $\pi$ bonding network of carbon atoms, which are $sp^2$ hybridized with three neighboring atoms.[18,37] On the other hand, in the case of defective samples, ferromagnetism has been recorded even at room temperature.[38,39] Especially, the large $\pi$ conjugation of residual oxygen-containing functional groups has been suggested as a source for the ferromagnetism in RGO.[40-43] Theoretical calculations have proposed that magnetic moments can be induced by forming a chemical bond of one carbon atom between two hydroxyl groups and a single C-C covalent bond established between an adsorbate and graphene.[40,44] In this sense, each chemical bonding electronic state can provide different magnetic signal. Therefore, based on our precise information on the chemical states of GO, RGO1, and RGO2 obtained from the above spectroscopic measurements, we may decompose their magnetic data with different chemical bonding state.

Figure 3(a) shows the temperature dependent magnetization $\chi(T)$ of GO, RGO1 and RGO2 measured in an applied magnetic field of 1 kOe after cooled at zero field at temperatures from 2 to 300 K. With decreasing temperature, $\chi(T)$ curve of GO initially increases and then increases rapidly at low temperatures, while $\chi(T)$ curves of RGO1 and RGO2 exhibit similar behavior with gradual rise except the difference of $\chi(T)$ values. According to the deconvoluted results of relative elemental composition percentage from XPS in Table 1, the amount of oxygen–containing functional groups, especially hydroxyl group, is the largest in GO, so that it can be a source for the highly temperature dependent $\chi(T)$ data. Upon reduction, in RGO1 and RGO2, the $sp^2$ bonding is recovered and the $\chi(T)$ curves become comparatively flat. These results give a hint that the temperature independent term such as Pauli paramagnetism and diamagnetism mainly contributes to the magnetism in RGO1 and RGO2. Moreover, the $\chi(T)$ values of RGO2 are negative above 7 K, as shown in the inset of Fig. 3(a), which is a clear indication of diamagnetism in RGO2.

The overall features of temperature dependent $\chi(T)$ are qualitatively analyzed using the modified Curie-Weiss law, $\chi-\chi_0 = C/(T-\theta_P)$, where $\chi_0$ represents the temperature-independent contribution, $C$ the Curie constant, and $\theta_P$ the paramagnetic Curie temperature. From the linear fit of $1/(\chi-\chi_0)$ vs. $T$ in Fig. 3(b), we obtain the parameters of $\chi_0$, $C$, and $\theta_P$, which are summarized in Table 2. Here note that the $1/T$ dependence of $\chi(T)$ represents the Langevin paramagnetism and the temperature independent contribution comes from the combination and/or competition of Pauli paramagnetism and diamagnetism. The largest $C$ value in GO suggests that the Langevin paramagnetism is the most dominant. Reminding that the compositional percentage of hydroxyl functional group is the highest in GO, we consider the Langevin paramagnetism in connection with the chemical bond of hydroxyl functional groups. This result is consistent with the previous report,[45] proposing that the hydroxyl functional groups induce robust magnetic moment on the basal plane of the grapheme sheet. On the other hand, the $C$ values are small and almost identical in both RGO1 and RGO2, while the $\chi_0$ values are large and different in sign. The $\chi_0$ value is the largest for RGO1 and is negative for RGO2. The large positive $\chi_0$ in RGO1 indicates more contribution of Pauli paramagnetism than diamagnetism. Since the compositional percentage of carboxyl functional groups in RGO1 is large as listed in Table

1, the origin of Pauli paramagnetism seems to be the carboxyl functional groups. This result agrees well with a recent theoretical study,[46] in which the carboxyl functional group in graphene sheets behaves as electron donors contributing to the Pauli paramagnetism of RGO1. For RGO2, since the compositional percentage of C=C($sp^2$) bonding is the largest, the negative $\chi_0$ is thought to be attributed to the restoration of $sp^2$ bonding.

For further understanding the magnetic behavior of each chemical functional group, we measured the magnetic field dependence of magnetization $M(H)$ for GO, RGO 1 and RGO 2 at different temperatures of 5 K and 300 K. The results are shown in Fig. 4. There is no hysteresis due to ferromagnetic origin in all the samples. For GO, since the Langevin paramagnetism caused by the hydroxyl groups is dominant, it is natural that $M(H)$ at 5 K shows a typical paramagnetic behavior with Brillouin function shape and it becomes diamagnetic at 300 K. For RGO1, the $M(H)$ curves are positively linear at 5 K and 300 K, which is ascribed to the temperature independent term of Pauli paramagnetism by donors generated from carboxyl groups. For RGO2, the diamagnetic $M(H)$ curves are observed at both temperatures except the low-field regime at 5 K. This result is close to the intact graphene, which should be nonmagnetic due to $\pi$-symmetry electronic states of $sp^2$ carbon atoms.

We explored the possibility that the magnetic data can be used as a means of determining the quality of RGO. For this purpose, we fabricated several graphene oxides of GO, RGO1, and RGO2 with different oxygen reduction, and carried out various spectroscopy measurements such as Raman, FT-IR, and XPS. Our spectroscopic results provided the precise information on the chemical states of RGO such as oxygen-containing functional groups, $sp^2$ carbon bonding, and reduction rate of GO. Recent theoretical calculations reported that each chemical bonding electronic state can induce different magnetic signal. Therefore, we investigated the relation between the chemical bonding state and the magnetic moment, and finally decomposed the magnetic data with different chemical bonding state. In conclusion, the magnetic data can serve as a new method to determine the quality of RGO in terms of mass production, although the spectroscopy data give the precise information on the chemical states of GO and RGO.

**Acknowledgements**

This work was supported by the National Research Foundation of Korea (NRF) grant funded by the Korea government (MEST) (No. 2014R1A2A1A1105401 and 2017R1A2B3007918).

Table 1. Percentage of Relative elemental composition of C=C ($sp^2$), C–C ($sp^3$), C–O (hydroxyl), C=O (carbonyl), and O–C=O (carboxyl) for GO, RGO1 and RGO2, estimated from the deconvolution of XPS C1s spectra.

|       | C=C   | C–C   | C–O   | C=O  | O–C=O | C/O  | C=C/C–C |
|-------|-------|-------|-------|------|-------|------|---------|
| GO    | 42.7  |       | 46.06 | 10.1 | 1.15  | 0.75 | -       |
| RGO 1 | 61.99 | 15.19 | 13.16 | 4.97 | 4.68  | 3.38 | 4.08    |
| RGO 2 | 69.41 | 15.03 | 9.89  | 3.32 | 2.35  | 5.43 | 4.62    |

Table 2. Fitting parameters of temperature-independent magnetic susceptibility $\chi_0$, Curie constant $C$, and paramagnetic Curie temperature $\theta_P$ for GO, RGO1 and RGO2, obtained from the Curie-Weiss law.

|       | $\chi_0$ ( $10^{-4}$ emu/g ) | $C$ ( emu K/g ) | $\theta_P$ ( K ) |
|-------|------------------------------|-----------------|------------------|
| GO    | 2.85                         | 0.098           | 0.13             |
| RGO 1 | 69.1                         | 0.017           | -0.03            |
| RGO 2 | -18.7                        | 0.018           | -0.21            |

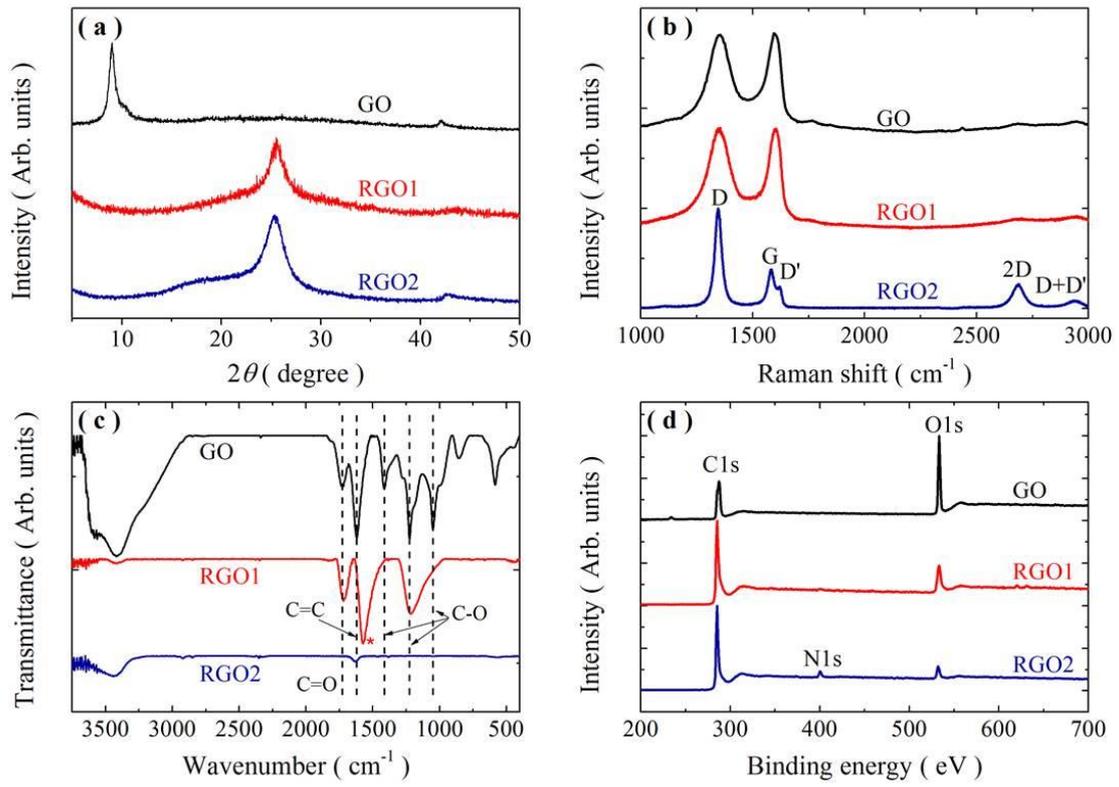

Fig. 1 (a) XRD patterns, (b) Raman spectra, (c) FT-IR spectra, and (d) wide-scan XPS spectra of GO, RGO1 and RGO2, respectively.

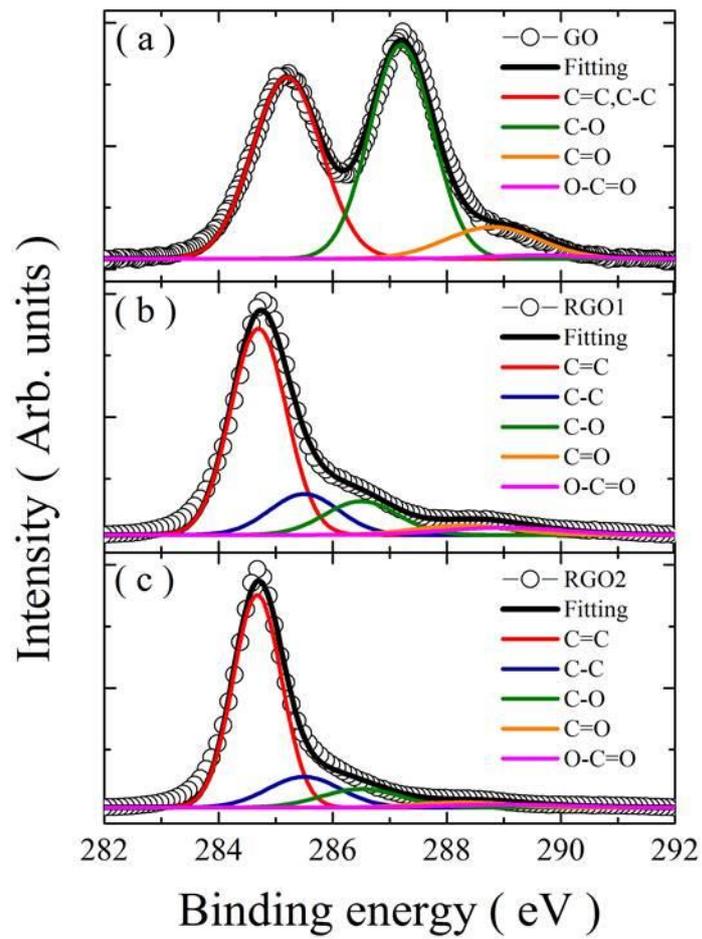

Fig. 2 Deconvoluted XPS C1s spectra of (a) GO, (b) RGO1, and (c) RGO2. The open circles represent the experimental data, and the black solid line display the sum of five components arising from C=C, C-C, C-O, C=O, and O-C=O carbon bonds.

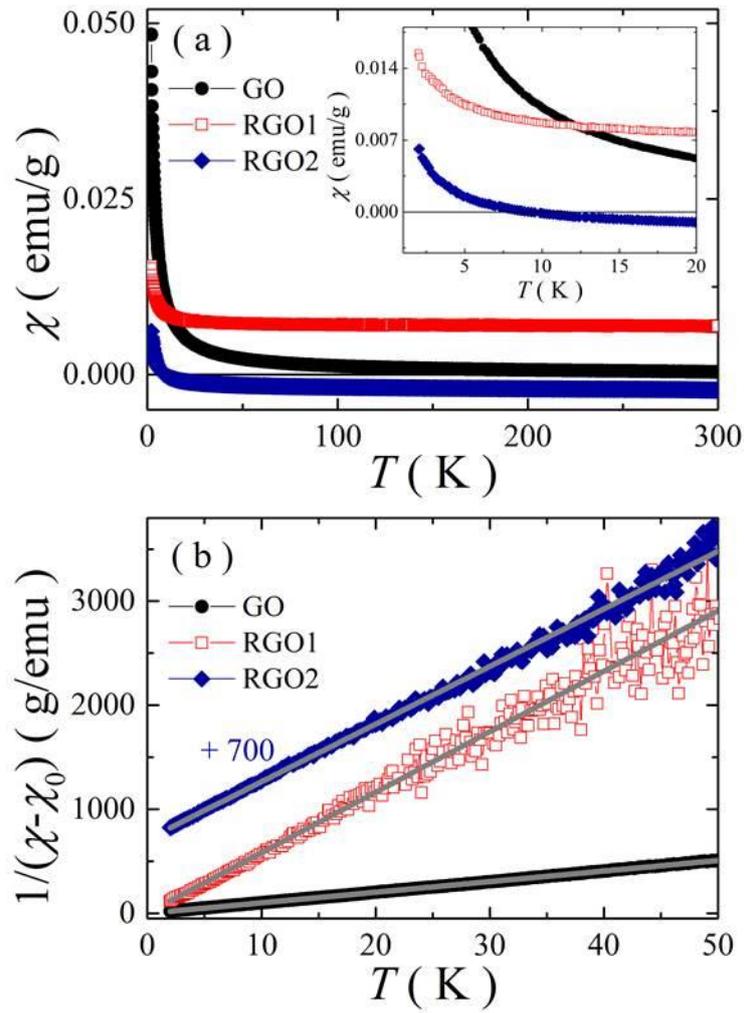

Fig. 3 (a) Temperature dependence of magnetization $\chi(T)$ for GO, RGO1, and RGO2 measured in an applied magnetic field of 1 kOe after zero-field cooling. The inset shows the magnified $\chi(T)$ in the low temperature region. (b) Temperature dependence of inverse magnetic susceptibility $1/(\chi-\chi_0)$ in the temperature range from 2 to 50 K, where the lines represent the linear fitting of the Curie-Weiss law. The data of RGO2 shifted to +700 g/emu in the plot.

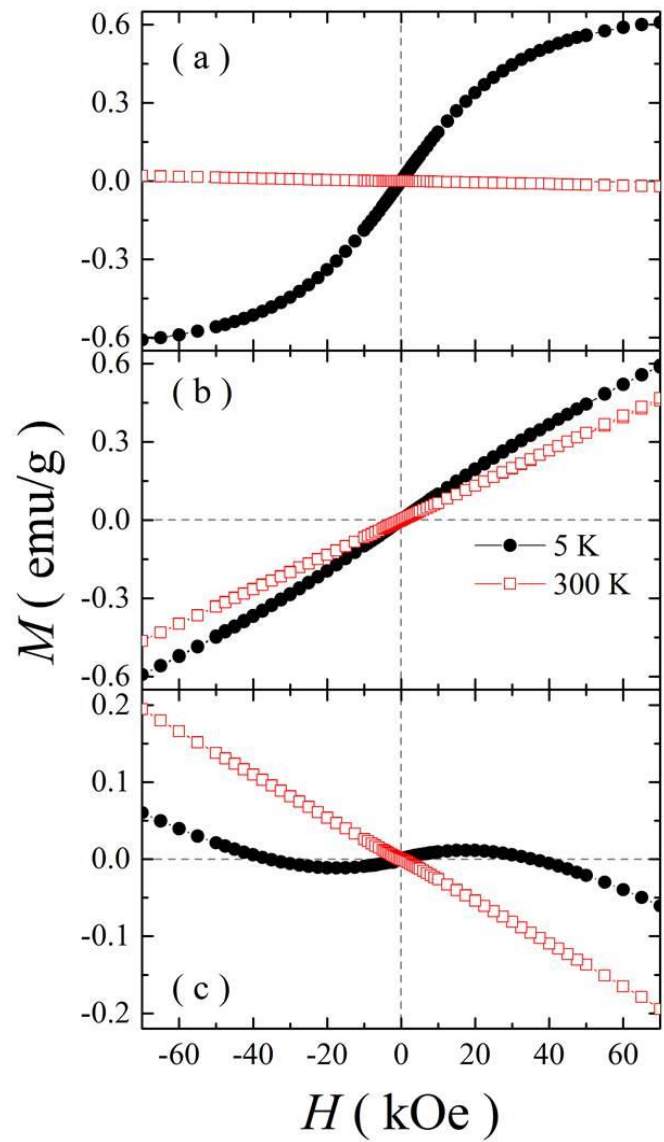

Fig. 4 Magnetic field dependence of magnetization $M(H)$ for (a) GO, (b) RGO1, and (c) RGO2 measured at 5 K and 300 K in the field range from -70 kOe to 70 kOe.